\documentclass{article}
\usepackage{spconf, amsmath,graphicx,hyperref}

\usepackage{amsfonts}
\usepackage{amssymb}
\usepackage{subcaption}
\usepackage{siunitx}
\usepackage{booktabs}

\title{Length-Aware Rotary Position Embedding for Text-Speech Alignment}
\name{Hyeongju Kim, Juheon Lee, Jinhyeok Yang, Jacob Morton\thanks{Corresponding author: \texttt{hyeongju@supertone.ai}}}
\address{Supertone, Inc.}
\begin{document}
\ninept
\maketitle
\begin{abstract}
Many recent text-to-speech (TTS) systems are built on transformer architectures and employ cross-attention mechanisms for text-speech alignment. Within these systems, rotary position embedding (RoPE) is commonly used to encode positional information in text and speech representations. In this work, we introduce length-aware RoPE (LARoPE), a simple yet effective extension of RoPE that improves text-speech alignment. Unlike RoPE, which relies on absolute indices, LARoPE computes relative distances between query and key positions using length-normalized indices. Experimental results show that LARoPE consistently outperforms RoPE, offering faster loss convergence, more accurate text-speech alignment, and higher overall TTS quality. Furthermore, LARoPE demonstrates greater resilience to variations in utterance duration and maintains stable performance in extended speech generation up to 30 seconds, whereas RoPE suffers from notable degradation. Notably, our method achieves a state-of-the-art word error rate on a standard zero-shot TTS benchmark.

\end{abstract}
\section{Introduction}
\label{sec:intro}

Optimization of generative models for text-to-speech (TTS) requires accurate alignments between text and speech. While some speech datasets provide phoneme-level timestamps, such detailed annotations are not available in most speech corpora~\cite{garofolo1993timit, zen2019libritts,librispeech_org, yamagishi2019vctk}. To address unknown text-speech alignments, several techniques have been explored in the TTS literature. One widely used approach is to employ a forced aligner during preprocessing to obtain explicit text-speech alignments~\cite{jeong21_interspeech}. 
Alternatively, a monotonic alignment search (MAS) algorithm can be integrated directly into the training process~\cite{kim2020glow,yang24q_interspeech}. However, both methods can introduce alignment errors, as they depend on the performance of a pretrained acoustic model or the model currently being trained. 

Recently, TTS models have increasingly moved toward using attention mechanisms for text conditioning~\cite{wang2023neural, kharitonov2023speak, kim2025supertonictts, lee2025dittotts,chen2024f5,eskimez2024e2}. These models implicitly infer text-speech alignments, eliminating the need for explicit alignment tools. Three primary approaches are commonly employed for this purpose. In the first approach, text embeddings are provided as prefix inputs to transformer blocks and interact with speech features through self-attention layers~\cite{wang2023neural, kharitonov2023speak}. In the second approach, text embeddings are conditioned on speech features via cross-attention layers, where the speech features serve as queries and the text embeddings serve as key and value vectors~\cite{kim2025supertonictts,lee2025dittotts}. In the third approach, text embeddings are padded with filler tokens to match the temporal length of speech features and concatenated with them along the channel axis~\cite{chen2024f5,eskimez2024e2}. Despite their flexibility, however, these attention-based models often suffer from errors such as repetitions, omissions, or mispronunciations due to the lack of explicit alignment supervision.



In this work, we aim to enhance \textbf{alignment learning} in TTS models that employ \textbf{cross-attention for text conditioning}. These models typically employ rotary position embeddings (RoPE)~\cite{su2024roformer} for both text and speech representations~\cite{kim2025supertonictts,lee2025dittotts}. However, we note that RoPE does not fully exploit the inherent monotonic relationship between text and speech, making it suboptimal for alignment learning. To this end, we propose \textbf{Length-Aware RoPE (LARoPE)}, a simple yet effective modification to RoPE. Unlike the original RoPE, which relies on absolute positional indices, LARoPE computes relative distances between query and key vectors using length-normalized indices. This adjustment induces a diagonal bias in the attention score maps, which better conforms to the monotonic structure of text-speech alignment.

Our experimental results demonstrate that LARoPE consistently outperforms RoPE across several key metrics. Specifically, LARoPE achieves faster loss convergence, reduced pronunciation errors, and better speaker similarity. It also exhibits robust performance for extended speech generation up to 30 seconds, whereas RoPE shows significant degradation. In addition, LARoPE is more resilient to variations in utterance duration. Finally, our proposed model achieves a state-of-the-art word error rate (WER) among zero-shot TTS models that rely on attention mechanisms for alignment learning. These findings confirm that LARoPE is a simple yet powerful positional embedding for robust, high-quality speech synthesis.


\section{Preliminary}
\label{section:preliminary}

RoPE~\cite{su2024roformer} is a sophisticated approach to positional encoding that effectively captures sequential information in transformer models. Unlike traditional positional embeddings which explicitly add position-dependent vectors to token representations, RoPE encodes positional information implicitly through a rotational transformation applied to the query and key vectors in the attention mechanism. 

Specifically, given a token embedding vector $x\in \mathbb{R}^{d}$ at position $p$, RoPE partitions $x$ into $d/2$ two-dimensional subvectors and applies a rotation to each subvector.\footnote{We assume that $d$ is an even number.} This process can be formally represented by the rotation operator $R_\theta(x,p)$ as follows:
\begin{equation}
R_\theta(x, p) = \begin{bmatrix}
    R_{\theta_0} & 0 & \cdots & 0 \\
    0 & R_{\theta_1} & \ddots & \vdots \\
    \vdots & \ddots & \ddots & 0 \\
    0 & \cdots & 0 & R_{\theta_{d/2-1}}
\end{bmatrix} x,
\end{equation}
where each rotation matrix $R_{\theta_j}\in\mathbb{R}^{2 \times 2}$ for the $j$-th subvector $x_{[2j:2j+1]}=\left[x_{2j}, x_{2j+1}\right]^\intercal\in \mathbb{R}^2$ performs a simple rotation:
\begin{equation}
R_{\theta_j}\begin{bmatrix}
x_{2j} \\
x_{2j+1}
\end{bmatrix}
=
\begin{bmatrix}
\cos(p\theta_{j}) & -\sin(p\theta_{j}) \\
\sin(p\theta_{j}) & \cos(p\theta_{j})
\end{bmatrix}
\begin{bmatrix}
x_{2j} \\
x_{2j+1}
\end{bmatrix}.
\end{equation}
The rotation frequency $\theta_{j}$ is typically defined as $\theta_{j}=10000^{-2j/d}$. 

In self-attention mechanism, for a query vector $q=W_qx^m$ at position $m$ and a key vector $k=W_k x^n$ at position $n$, RoPE applies the corresponding rotation operators $R_\theta(\cdot, m)$ and $R_\theta(\cdot, n)$, respectively, and computes their inner product $ R_\theta(q, m)^\intercal R_\theta(k, n)$ as
\begin{equation}
\label{eq:rope_inner_product}
\text{Re}\left[\sum_{j=0}^{d/2-1} q_{[2j:2j+1]} k^*_{[2j:2j+1]} e^{i(m-n)\theta_j} \right],
\end{equation}
where $i$ denotes the imaginary unit and $^*$ indicates complex conjugation. This formulation makes the attention score depend only on the relative positional difference $(m-n)$, thereby naturally encoding relative positional relationships. Furthermore, the upper bound of Eq.~\ref{eq:rope_inner_product} can be expressed as
\begin{equation}
\label{eq:rope_upper_bound}
\left(\max_j|h_{j+1} - h_j|\right)\sum_{j=0}^{d/2-1}S_j,
\end{equation}
where $h_j = q_{[2j:2j+1]} k^*_{[2j:2j+1]}$ and $S_j=\sum_{k=0}^j\left|e^{i(m-n)\theta_k}\right|$. This bound reveals the long-range decay property of RoPE, which helps the model generalize across varying sequence lengths and improves extrapolation to longer contexts.
\begin{figure}[t]
    \centering
    \includegraphics[width=0.95\columnwidth]{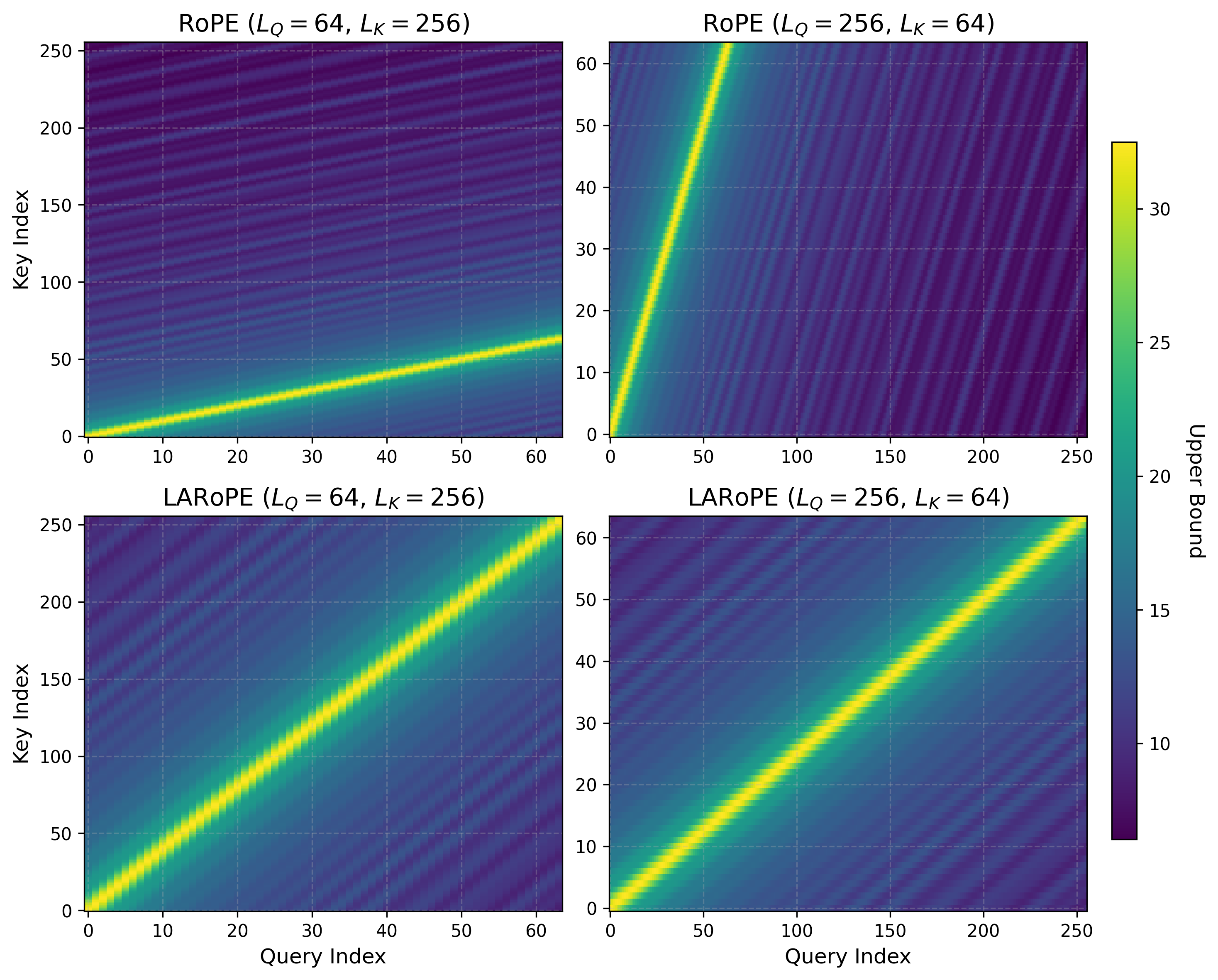}
    \caption{Comparison of relative upper bounds with different combinations of query length ($L_q$) and key length ($L_k$).}
    \label{fig:ub1}
\end{figure}
\section{Proposed Method}
\label{section:proposed_method}

Contemporary TTS models are usually built on transformer blocks and employ cross-attention layers to condition speech features on textual representations~\cite{kim2025supertonictts, lee2025dittotts}. In this setup, speech features act as queries, while text embeddings serve as keys. To incorporate positional information, RoPE is typically applied to both queries and keys before computing the attention scores. However, unlike in self-attention, the relative positional offsets introduced by RoPE are no longer aligned along a diagonal, since the query and key sequences generally differ in length. Note that the relative distance in Eq.~\ref{eq:rope_inner_product} is evaluated with respect to the absolute positional indices of query and key vectors. Consequently, RoPE in cross-attention deviates from the diagonal-like alignment that would naturally be expected between text and speech, making it less effective for text conditioning.

To address this issue, we introduce LARoPE, a method for encoding positional information that adapts to varying sequence lengths. LARoPE modifies the rotation matrix $R_{\theta_j}$ by incorporating the input length. Specifically, for an input vector $x$ at position $p$ in a sequence of length $L$, the length-aware rotation operator $R'_{\theta}(x,p,L)$ is defined as follows:
\begin{equation}
\label{eq:LARoPE_rotation_operator}
R'_\theta(x, p,L) = 
\begin{bmatrix}
    R'_{ \theta_0} & 0 & \cdots & 0 \\
    0 & R'_{ \theta_1} & \ddots & \vdots \\
    \vdots & \ddots & \ddots & 0 \\
    0 & \cdots & 0 & R'_{ \theta_{d/2-1}}
\end{bmatrix} x,
\end{equation}
where each rotation matrix $R'_{\theta_j}\in\mathbb{R}^{2 \times 2}$ is given by
\begin{equation}
\label{eq:LARoPE_rotation_matrix}
R'_{\theta_j}
=
\begin{bmatrix}
\cos\!\left(\gamma \frac{p}{L}\theta_{j}\right) & -\sin\!\left(\gamma \frac{p}{L}\theta_{j}\right) \\[1ex]
\sin\!\left(\gamma \frac{p}{L}\theta_{j}\right) & \cos\!\left(\gamma \frac{p}{L}\theta_{j}\right)
\end{bmatrix},
\end{equation}
with $\gamma$ denoting a scaling hyperparameter. 

In a cross-attention layer where the query length $L_q$ and the key length $L_k$ generally differ, the inner product between $R'_\theta(q,m,L_q)$ and $R'_\theta(k,n,L_k)$ can be expressed as:
\begin{equation}
\label{eq:larope_inner_product}
\text{Re}\left[\sum_{j=0}^{d/2-1} q_{[2j:2j+1]} k^*_{[2j:2j+1]} e^{i\gamma(\frac{m}{L_q}-\frac{n}{L_k})\theta_j} \right].
\end{equation}
Since Eq.~\ref{eq:larope_inner_product} calculates relative distances based on length-normalized indices, its upper bound remains aligned along a diagonal even when query and key lengths are different. 

To understand the impact on the attention score map, we analyze the relative upper bound by calculating $\sum_{j=0}^{d/2-1}S_j$. In the case of LARoPE, each term $S_j$ is given by $\sum_{k=0}^j\left|e^{i\gamma(\frac{m}{L_q}-\frac{n}{L_k})\theta_j}\right|$. We specifically examine two pairs of query and key lengths: $(L_q, L_k)=(64, 256)$ and $(256, 64)$.  As illustrated in Fig.~\ref{fig:ub1}, LARoPE consistently preserves a diagonal structure in the relative upper bound regardless of sequence length, unlike RoPE. 
This diagonal property is analogous to the natural alignment between text and speech, thereby facilitating more effective alignment learning in cross-attention.

\section{Experiments}
\label{section:experiments}

\subsection{Experimental setup}

We evaluate the effectiveness of LARoPE within the SupertonicTTS framework~\cite{kim2025supertonictts}. SupertonicTTS introduces batch expansion to accelerate text-speech alignment. However, since our primary goal is to assess the impact of LARoPE on alignment itself, we also consider a baseline model without batch expansion (i.e., expansion factor $K_e=1$) to simulate a more challenging alignment scenario. Specifically, we compare four text-to-latent modules in SupertonicTTS by varying both the positional embedding (RoPE or LARoPE) and the expansion factor ($K_e=1$ or $K_e=4$). 

For training, we used a composite of high-quality TTS datasets, including LJSpeech~\cite{ljspeech17}, VCTK~\cite{yamagishi2019vctk}, Hi-Fi TTS~\cite{bakhturina2021hi}, and LibriTTS~\cite{zen2019libritts} datasets. The models were optimized using AdamW~\cite{loshchilov2018decoupled} for 700k training iterations with a batch size of 64 on four NVIDIA RTX 4090 GPUs. The initial learning rate was set to {\num{5e-4}} and reduced by half every 300k iterations. For LARoPE, the scaling parameter $\gamma$ was set to 10. All audio files were resampled to {\SI{44.1}{kHz}} when their original sampling rate differed. 

For evaluation, we set the number of function evaluations (NFE) to 32 and the classifier-free guidance scale to 3. Model outputs were transcribed using the CTC-based HuBERT-large ASR model\footnote{\href{https://huggingface.co/facebook/hubert-large-ls960-ft}{https://huggingface.co/facebook/hubert-large-ls960-ft}.} and normalized~\cite{zhang2021nemo} to compute character error rate (CER) and word error rate (WER). Speaker embeddings were extracted with the WavLM-TDNN model~\cite{chen2022wavlm}, and cosine distance between generated and reference speech was used to measure speaker similarity (SIM). 
Finally, an objective score for perceptual quality was obtained using the UTMOSv2 model\cite{baba2024utmosv2}, which serves as a proxy for human mean opinion score (MOS) ratings.


\subsection{Performance evaluation}
\label{subsection:performance_comparison_librispeech}
\begin{table}[]
\centering
\caption{Performance comparison on LibriSpeech test-clean subsets. ``Pos.~emb.'' refers to positional embedding.}
\begin{tabular}{lcccc}
\toprule
Pos. emb. & ${K_e}$ & WER(\%) & SIM & UTMOSv2 \\ \midrule
\multicolumn{5}{c}{\textit{tc-short} (4 - 10 sec.)}\\ \midrule
RoPE & 1 & 2.62& 0.46~\scriptsize{± 0.01} & 3.05~\scriptsize{± 0.01} \\
LARoPE & 1 & \textbf{2.34} & \textbf{0.49~\scriptsize{± 0.01}} & \textbf{3.11~\scriptsize{± 0.01}} \\ \midrule
RoPE & 4 & 2.41& 0.48~\scriptsize{± 0.01} & \textbf{3.23~\scriptsize{± 0.01}} \\
LARoPE & 4 & \textbf{2.25} & \textbf{0.50~\scriptsize{± 0.01}} & 3.21~\scriptsize{± 0.01} \\ \midrule
\multicolumn{5}{c}{\textit{tc-long} (10 - 30 sec.)} \\ \midrule
RoPE & 1 & 4.87& 0.50~\scriptsize{± 0.01} & 3.08~\scriptsize{± 0.01} \\
LARoPE & 1 & \textbf{2.24} & \textbf{0.52~\scriptsize{± 0.01}} & \textbf{3.21~\scriptsize{± 0.01}} \\ \midrule
RoPE & 4 & 4.98& 0.52~\scriptsize{± 0.01} & \textbf{3.27~\scriptsize{± 0.01}} \\
LARoPE & 4 & \textbf{2.16} & \textbf{0.54~\scriptsize{± 0.01}} & \textbf{3.27~\scriptsize{± 0.01}} \\ \bottomrule
\end{tabular}
\label{table:librispeech}
\end{table}

\begin{table}[]
\caption{WER(\%) comparison across different duration scaling factors, where $d$ denotes the original utterance duration.}
\centering
\begin{tabular}{l|ccccc}
\toprule
                 & $0.7d$       & $0.85d$       & $1.0d$        & $1.2d$        & $1.4d$        \\ \midrule
RoPE ($K_e=4$)   & 6.03          & 2.61          & 2.41          & 3.66          & 7.39          \\
LARoPE ($K_e=4$) & \textbf{4.77} & \textbf{2.48} & \textbf{2.25} & \textbf{2.98} & \textbf{5.40} \\ \bottomrule
\end{tabular}
\label{table:duration}
\end{table}

We evaluated model performance using two subsets derived from the test-clean split of LibriSpeech~\cite{librispeech_org}: \textit{tc-short} and \textit{tc-long}. The \textit{tc-short} set, which is identical to the test set introduced by \cite{chen2024f5}, consists of 1,127 utterances between 4 and 10 seconds in duration. The \textit{tc-long} set includes longer utterances, ranging from 10 to 30 seconds. Following \cite{kim2025supertonictts}, we used the transcriptions from LibriSpeech-PC~\cite{meister2023librispeech} as model input when available; otherwise, we capitalized the original LibriSpeech transcription and appended a period.

The evaluation results are shown in Table~\ref{table:librispeech}. Across both subsets and both $K_e$ values, LARoPE consistently outperforms RoPE in terms of WER and SIM, demonstrating its effectiveness in correct pronunciation and speaker similarity. Specifically, LARoPE achieves notable reductions in WER, especially on the \textit{tc-long} subset (e.g., 4.98 → 2.16 when $K_e=4$), showing its robustness in generating longer utterances. In terms of speaker similarity, LARoPE consistently yields higher scores than RoPE across all conditions, indicating that it also helps to capture speaker characteristics. For perceptual quality estimated by UTMOSv2, the gains are also evident when $K_e=1$ (e.g., 3.08 → 3.21 on the \textit{tc-long} set). Overall, these results provide clear evidence of LARoPE’s advantages over RoPE.

We further investigate the robustness of LARoPE to variations in utterance duration, which can occur when users wish to synthesize speech faster or slower than the original pace. Using the $\textit{tc-short}$ set, we applied different duration scaling factors of 0.7, 0.85, 1.0, 1.2, and 1.4, and measured the resulting WERs, as summarized in Table~\ref{table:duration}. Across all scaling factors, LARoPE consistently outperforms RoPE for both shortened and lengthened utterances. These results indicate that LARoPE is more robust to temporal variations and better preserves intelligibility under diverse speech rates. 


\subsection{Performance evolution across training}
\begin{figure}[t]
    \centering
    \includegraphics[width=1.0\columnwidth]{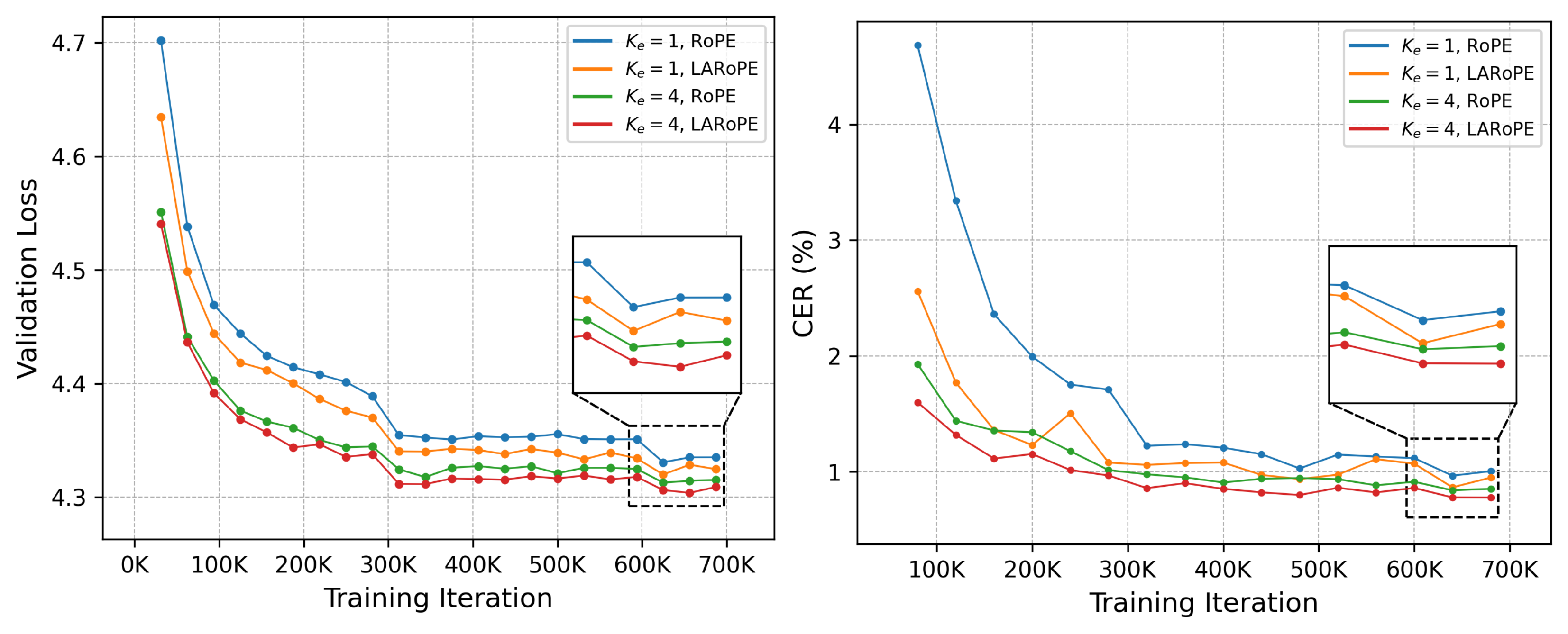}
    \caption{Evolution of validation loss and CER curves during training.}
    \label{fig:curve}
\end{figure}
    



LARoPE is designed with a diagonal structure in its relative upper bound, which we hypothesize to be more effective for text-speech alignment learning compared to RoPE. To validate this, we tracked the flow matching losses and CERs over the course of training.

As shown in Fig.~\ref{fig:curve}, LARoPE consistently outperforms RoPE in both validation loss and CER. The advantage is particularly notable when batch expansion is disabled ($K_e=1$), a setting where alignment learning becomes more challenging. At iteration 200k, LARoPE reduces CER from 2.00\% to 1.23\% in this setting, highlighting its efficiency in facilitating alignment. A similar trend is observed with batch expansion ($K_e = 4$) at the same iteration, where LARoPE achieves 1.15\% CER compared to 1.34\% for RoPE. This performance gap persists throughout training, leading to improved final accuracy. Overall, these results indicate that LARoPE serves as a more effective positional embedding for TTS, accelerating the alignment learning process.


\subsection{Analysis of attention score map}
\begin{figure}[t]
    \centering
    \includegraphics[width=1.0\columnwidth]{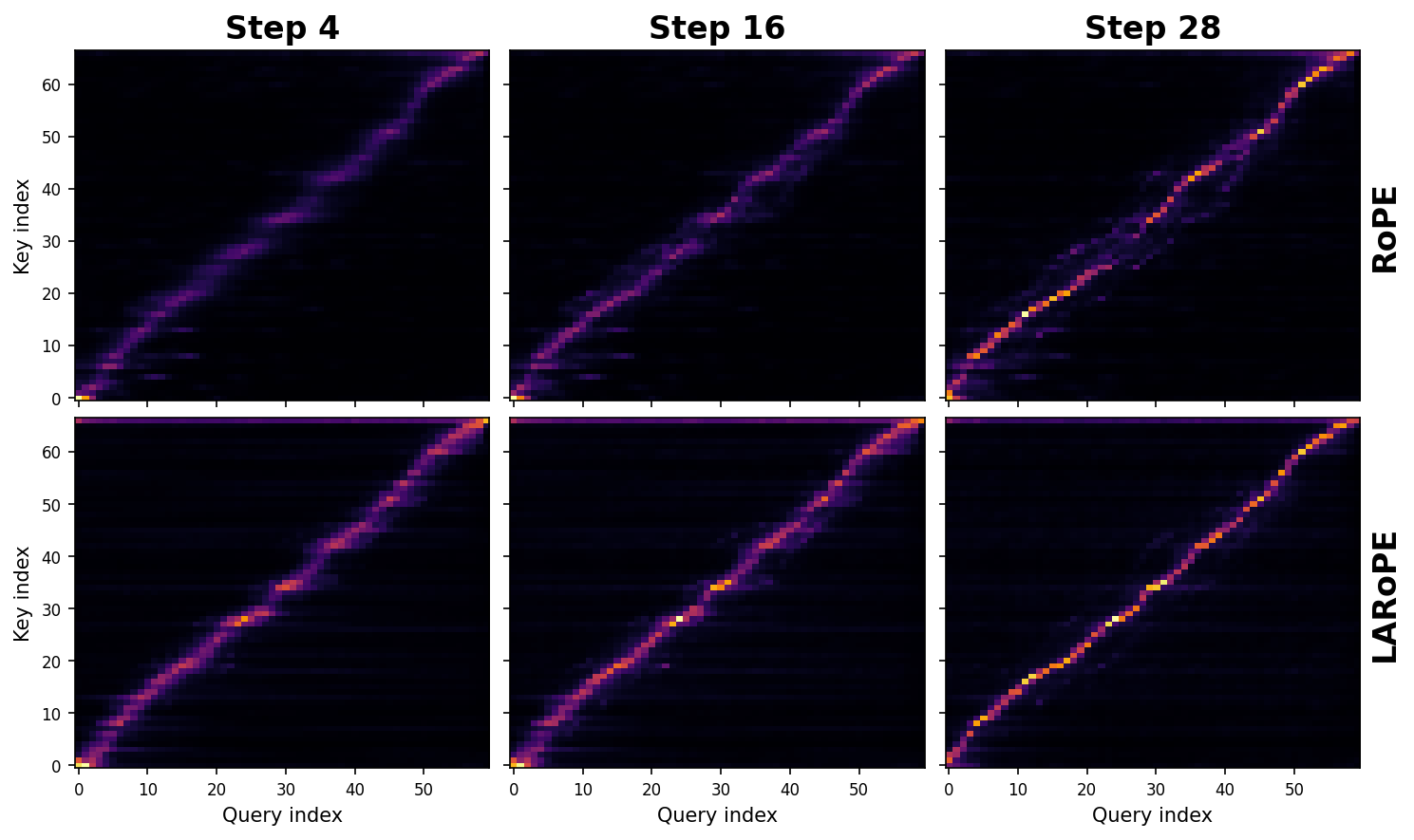}
    \caption{Comparison of average attention score map.}
    \label{fig:attention_score}
\end{figure}
To investigate the effect of LARoPE on attention mechanisms, we compared the average attention score maps generated by RoPE and LARoPE. Specifically, we collected all attention score maps from the text-conditioning cross-attention layers and averaged them into a single representative map. For better interpretability, we excluded indices corresponding to space characters in the key positions, which consistently exhibited disproportionately high scores.\footnote{We hypothesize that space characters are mainly used for prosody modeling, such as pauses and word duration, rather than phonetic alignment.}

A representative example is shown in Fig.~\ref{fig:attention_score}. Overall, the attention maps from LARoPE are clearer and more continuous compared to those from RoPE. In particular, LARoPE produces sharper and more concentrated attention distributions in the early inference steps (steps 4 and 16). We attribute this behavior to LARoPE's diagonal bias, which strengthens text-speech alignment. This improved alignment leads to more stable and coherent attention score maps in later steps (e.g., step 28). These observations provide further empirical evidence for the effectiveness of LARoPE as a positional embedding in text-to-speech models.


\subsection{Comparison with other TTS models}
\begin{table}[t]
\caption{Comparison with other TTS models. ``\#Param.'' indicates the total number of learnable parameters in each model, excluding those of the vocoder. Results with $^*$ are obtained with the test-clean subset of LibriSpeech.}
\centering
\begin{tabular}{@{}lcccc@{}}
\toprule
              & Pos. emb. & \#Param. & RTF           & WER(\%)       \\ \midrule
Ground-Truth  & -         & -        & -             & 1.86          \\ \midrule
DiTTo-TTS     & RoPE      & 740M     & -            & 2.56$^*$      \\
E2 TTS        & RoPE      & 333M     & 0.68          & 2.95          \\
F5-TTS        & RoPE      & 336M     & 0.31          & 2.42          \\
SupertonicTTS & RoPE      & 19M      & \textbf{0.05} & 2.41          \\
Ours          & LARoPE    & 19M      & \textbf{0.05} & \textbf{2.25} \\ \bottomrule
\end{tabular}
\label{table:other_tts}
\end{table}

We benchmark our model against several state-of-the-art zero-shot TTS models that employ RoPE for positional embeddings. The selected baselines include E2 TTS~\cite{eskimez2024e2}, F5-TTS~\cite{chen2024f5}, DiTTo-TTS~\cite{lee2025dittotts}, and SupertonicTTS~\cite{kim2025supertonictts}. The WER results for DiTTo-TTS are obtained using the 2.2-hour test-clean subset of LibriSpeech, while results for the other models are evaluated on the \textit{tc-short} set used in Section~\ref{subsection:performance_comparison_librispeech}. Also, real-time factor (RTF) is measured for a 10-second audio generation using an NVIDIA RTX 3090 GPU. For E2 TTS, we present the results obtained from the reproduced model in \cite{chen2024f5}.

The comparative results are presented in Table~\ref{table:other_tts}. Our proposed model achieves a WER of 2.25\%, which surpasses all baseline models. Remarkably, it maintains the same parameter count (19M) and RTF (0.05) as SupertonicTTS, demonstrating that LARoPE enhances pronunciation accuracy without incurring additional computational cost. In contrast, larger models such as DiTTo-TTS with 740M parameters and F5-TTS with 336M parameters exhibit higher WER or slower inference, highlighting the efficiency and effectiveness of our approach.

\section{Related work}
\label{section:related_works}

\subsection{Rotary position embedding}
RoPE encodes relative positions by rotating query and key vectors and has become the default positional encoding in transformer-based speech and language models~\cite{su2024roformer, kim2025supertonictts, lee2025dittotts}. Beyond the original formulation, several extensions such as position interpolation~\cite{chen2023extending}, YaRN~\cite{peng2023yarn}, and LongRoPE~\cite{ding2024longrope} have been proposed to improve stability and extend RoPE to very long contexts. However, these approaches primarily focus on self-attention extrapolation and do not address the challenge of cross-attention alignment when query and key sequences differ in length. Our work tackles this gap by inducing a monotonic bias on cross-attention score maps, specifically designed to improve the alignment between text and speech representations.

\subsection{Strategy for improved text-speech alignment}
Aligning text and speech representations remains a critical challenge in TTS, particularly for long-form speech or previously unseen text. Misalignment is a primary source of common synthesis errors such as repetitions, skips, and mispronunciations. Prior work has addressed this problem using various strategies. For instance, guided attention loss accelerates the training of attention modules by encouraging near-diagonal attention patterns~\cite{tachibana2018efficiently}. Another approach is to incorporate diagonal priors during monotonic alignment search~\cite{shih2021rad, badlani2022one}. However, these methods often rely on external aligners or require carefully tuned auxiliary loss terms. In contrast, our proposed method, LARoPE, provides a more streamlined solution. It requires no auxiliary losses, modified prior distributions, or external aligners. Instead, LARoPE can be seamlessly integrated into existing cross-attention layers, maintaining the original model size and training/inference cost. 

\section{Conclusion}
\label{section:conclusion}
We introduced LARoPE, a method designed to enhance text-speech alignment in TTS models that rely on cross-attention for text conditioning. As a length-aware extension of rotary position embeddings, LARoPE normalizes positional indices by sequence length when computing relative distances between query and key vectors. This strategy induces a diagonal bias in the attention score map, thereby leading to more stable and accurate alignments. Extensive experiments demonstrate consistent improvements over RoPE in word error rate, speaker similarity, and perceptual quality. Notably, LARoPE offers a significant advantage for extended speech generation, maintaining stable performance up to 30 seconds where RoPE degrades. Moreover, LARoPE is more resilient to variations in utterance duration. Analysis of average attention score maps further confirms that LARoPE encourages proper alignments in the early inference steps, resulting in clearer and more coherent attention patterns throughout the inference process. When compared with state-of-the-art zero-shot TTS systems, our model achieves the lowest WER while maintaining a compact architecture and low inference cost. These results highlight the importance of positional embedding design and suggest that LARoPE can serve as a simple and effective solution for robust, high-quality speech synthesis.

\pagebreak

\bibliographystyle{IEEEbib}
\bibliography{refs}

\begin{thebibliography}{10}

\bibitem{garofolo1993timit}
John~S Garofolo,
\newblock ``Timit acoustic phonetic continuous speech corpus,''
\newblock {\em Linguistic Data Consortium, 1993}, 1993.

\bibitem{zen2019libritts}
Heiga Zen, Viet Dang, Rob Clark, Yu~Zhang, Ron~J Weiss, Ye~Jia, Zhifeng Chen, and Yonghui Wu,
\newblock ``Libritts: A corpus derived from librispeech for text-to-speech,''
\newblock {\em arXiv preprint arXiv:1904.02882}, 2019.

\bibitem{librispeech_org}
Vassil Panayotov, Guoguo Chen, Daniel Povey, and Sanjeev Khudanpur,
\newblock ``Librispeech: An asr corpus based on public domain audio books,''
\newblock in {\em 2015 IEEE International Conference on Acoustics, Speech and Signal Processing (ICASSP)}, 2015, pp. 5206--5210.

\bibitem{yamagishi2019vctk}
Junichi Yamagishi, Christophe Veaux, and Kirsten MacDonald,
\newblock ``{CSTR VCTK Corpus}: English multi-speaker corpus for {CSTR} voice cloning toolkit (version 0.92),'' 2019.

\bibitem{jeong21_interspeech}
Myeonghun Jeong, Hyeongju Kim, Sung~Jun Cheon, Byoung~Jin Choi, and Nam~Soo Kim,
\newblock ``Diff-tts: A denoising diffusion model for text-to-speech,''
\newblock in {\em Interspeech 2021}, 2021, pp. 3605--3609.

\bibitem{kim2020glow}
Jaehyeon Kim, Sungwon Kim, Jungil Kong, and Sungroh Yoon,
\newblock ``Glow-tts: A generative flow for text-to-speech via monotonic alignment search,''
\newblock {\em Advances in Neural Information Processing Systems}, vol. 33, pp. 8067--8077, 2020.

\bibitem{yang24q_interspeech}
Jinhyeok Yang, Junhyeok Lee, Hyeong-Seok Choi, Seunghoon Ji, Hyeongju Kim, and Juheon Lee,
\newblock ``{DualSpeech: Enhancing Speaker-Fidelity and Text-Intelligibility Through Dual Classifier-Free Guidance},''
\newblock in {\em {Interspeech 2024}}, 2024, pp. 4423--4427.

\bibitem{wang2023neural}
Chengyi Wang, Sanyuan Chen, Yu~Wu, Ziqiang Zhang, Long Zhou, Shujie Liu, Zhuo Chen, Yanqing Liu, Huaming Wang, Jinyu Li, et~al.,
\newblock ``Neural codec language models are zero-shot text to speech synthesizers,''
\newblock {\em arXiv preprint arXiv:2301.02111}, 2023.

\bibitem{kharitonov2023speak}
Eugene Kharitonov, Damien Vincent, Zal{\'a}n Borsos, Rapha{\"e}l Marinier, Sertan Girgin, Olivier Pietquin, Matt Sharifi, Marco Tagliasacchi, and Neil Zeghidour,
\newblock ``Speak, read and prompt: High-fidelity text-to-speech with minimal supervision,''
\newblock {\em Transactions of the Association for Computational Linguistics}, vol. 11, pp. 1703--1718, 2023.

\bibitem{kim2025supertonictts}
Hyeongju Kim, Jinhyeok Yang, Yechan Yu, Seunghun Ji, Jacob Morton, Frederik Bous, Joon Byun, and Juheon Lee,
\newblock ``Supertonictts: Towards highly scalable and efficient text-to-speech system,''
\newblock {\em arXiv preprint arXiv:2503.23108}, 2025.

\bibitem{lee2025dittotts}
Keon Lee, Dong~Won Kim, Jaehyeon Kim, Seungjun Chung, and Jaewoong Cho,
\newblock ``Di{TT}o-{TTS}: Diffusion transformers for scalable text-to-speech without domain-specific factors,''
\newblock in {\em The Thirteenth International Conference on Learning Representations}, 2025.

\bibitem{chen2024f5}
Yushen Chen, Zhikang Niu, Ziyang Ma, Keqi Deng, Chunhui Wang, Jian Zhao, Kai Yu, and Xie Chen,
\newblock ``F5-tts: A fairytaler that fakes fluent and faithful speech with flow matching,''
\newblock {\em arXiv preprint arXiv:2410.06885}, 2024.

\bibitem{eskimez2024e2}
Sefik~Emre Eskimez, Xiaofei Wang, Manthan Thakker, Canrun Li, Chung-Hsien Tsai, Zhen Xiao, Hemin Yang, Zirun Zhu, Min Tang, Xu~Tan, et~al.,
\newblock ``E2 tts: Embarrassingly easy fully non-autoregressive zero-shot tts,''
\newblock in {\em 2024 IEEE Spoken Language Technology Workshop (SLT)}. IEEE, 2024, pp. 682--689.

\bibitem{su2024roformer}
Jianlin Su, Murtadha Ahmed, Yu~Lu, Shengfeng Pan, Wen Bo, and Yunfeng Liu,
\newblock ``Roformer: Enhanced transformer with rotary position embedding,''
\newblock {\em Neurocomputing}, vol. 568, pp. 127063, 2024.

\bibitem{ljspeech17}
Keith Ito and Linda Johnson,
\newblock ``The lj speech dataset,'' \url{https://keithito.com/LJ-Speech-Dataset/}, 2017.

\bibitem{bakhturina2021hi}
Evelina Bakhturina, Vitaly Lavrukhin, Boris Ginsburg, and Yang Zhang,
\newblock ``{Hi-Fi Multi-Speaker English TTS Dataset},''
\newblock {\em arXiv preprint arXiv:2104.01497}, 2021.

\bibitem{loshchilov2018decoupled}
Ilya Loshchilov and Frank Hutter,
\newblock ``Decoupled weight decay regularization,''
\newblock in {\em International Conference on Learning Representations}, 2019.

\bibitem{zhang2021nemo}
Yang Zhang, Evelina Bakhturina, Kyle Gorman, and Boris Ginsburg,
\newblock ``Nemo inverse text normalization: From development to production,''
\newblock {\em arXiv preprint arXiv:2104.05055}, 2021.

\bibitem{chen2022wavlm}
Sanyuan Chen, Chengyi Wang, Zhengyang Chen, Yu~Wu, Shujie Liu, Zhuo Chen, Jinyu Li, Naoyuki Kanda, Takuya Yoshioka, Xiong Xiao, et~al.,
\newblock ``Wavlm: Large-scale self-supervised pre-training for full stack speech processing,''
\newblock {\em IEEE Journal of Selected Topics in Signal Processing}, vol. 16, no. 6, pp. 1505--1518, 2022.

\bibitem{baba2024utmosv2}
Kaito Baba, Wataru Nakata, Yuki Saito, and Hiroshi Saruwatari,
\newblock ``The t05 system for the {V}oice{MOS} {C}hallenge 2024: Transfer learning from deep image classifier to naturalness {MOS} prediction of high-quality synthetic speech,''
\newblock in {\em IEEE Spoken Language Technology Workshop (SLT)}, 2024.

\bibitem{meister2023librispeech}
Aleksandr Meister, Matvei Novikov, Nikolay Karpov, Evelina Bakhturina, Vitaly Lavrukhin, and Boris Ginsburg,
\newblock ``Librispeech-pc: Benchmark for evaluation of punctuation and capitalization capabilities of end-to-end asr models,''
\newblock in {\em 2023 IEEE automatic speech recognition and understanding workshop (ASRU)}. IEEE, 2023, pp. 1--7.

\bibitem{chen2023extending}
Shouyuan Chen, Sherman Wong, Liangjian Chen, and Yuandong Tian,
\newblock ``Extending context window of large language models via positional interpolation,''
\newblock {\em arXiv preprint arXiv:2306.15595}, 2023.

\bibitem{peng2023yarn}
Bowen Peng, Jeffrey Quesnelle, Honglu Fan, and Enrico Shippole,
\newblock ``Yarn: Efficient context window extension of large language models,''
\newblock {\em arXiv preprint arXiv:2309.00071}, 2023.

\bibitem{ding2024longrope}
Yiran Ding, Li~Lyna Zhang, Chengruidong Zhang, Yuanyuan Xu, Ning Shang, Jiahang Xu, Fan Yang, and Mao Yang,
\newblock ``Longro{PE}: Extending {LLM} context window beyond 2 million tokens,''
\newblock in {\em Forty-first International Conference on Machine Learning}, 2024.

\bibitem{tachibana2018efficiently}
Hideyuki Tachibana, Katsuya Uenoyama, and Shunsuke Aihara,
\newblock ``Efficiently trainable text-to-speech system based on deep convolutional networks with guided attention,''
\newblock in {\em 2018 IEEE international conference on acoustics, speech and signal processing (ICASSP)}. IEEE, 2018, pp. 4784--4788.

\bibitem{shih2021rad}
Kevin~J Shih, Rafael Valle, Rohan Badlani, Adrian Lancucki, Wei Ping, and Bryan Catanzaro,
\newblock ``Rad-tts: Parallel flow-based tts with robust alignment learning and diverse synthesis,''
\newblock in {\em ICML Workshop on Invertible Neural Networks, Normalizing Flows, and Explicit Likelihood Models}, 2021.

\bibitem{badlani2022one}
Rohan Badlani, Adrian {\L}a{\'n}cucki, Kevin~J Shih, Rafael Valle, Wei Ping, and Bryan Catanzaro,
\newblock ``One tts alignment to rule them all,''
\newblock in {\em ICASSP 2022-2022 IEEE International Conference on Acoustics, Speech and Signal Processing (ICASSP)}. IEEE, 2022, pp. 6092--6096.

\end{thebibliography}

\end{document}